\documentclass[prl,twocolumn,superscriptaddress,showpacs]{revtex4} 
  
\usepackage[german,english]{babel} 
\usepackage{amssymb} 
\usepackage{amsfonts} 
 
\begin{document} 
 
\title{Equilibrium shapes of flat knots}
 
\author{Ralf Metzler} 
\affiliation{Department of Physics, Massachusetts Institute of Technology, 
Cambridge, MA 02139, USA} 
\author{Andreas Hanke} 
\affiliation{Department of Physics, Massachusetts Institute of Technology, 
Cambridge, MA 02139, USA} 
\author{Paul G. Dommersnes} 
\affiliation{Department of Physics, Massachusetts Institute of Technology, 
Cambridge, MA 02139, USA} 
\author{Yacov Kantor} 
\affiliation{School of Physics and Astronomy, Sackler Faculty of Exact 
Sciences, Tel Aviv University, Tel Aviv 69978, Israel} 
\affiliation{Department of Physics, Massachusetts Institute of Technology, 
Cambridge, MA 02139, USA} 
\author{Mehran Kardar} 
\affiliation{Department of Physics, Massachusetts Institute of Technology, 
Cambridge, MA 02139, USA} 
\affiliation{Institute for Theoretical Physics, University of California 
at Santa Barbara, Santa Barbara, CA 93106} 
 
\date{\today}  
 
\begin{abstract} 
We study the equilibrium shapes of prime and composite knots
confined to two dimensions. Using rigorous scaling arguments 
we show that, due to self-avoiding effects, the topological 
details of prime knots are localised on a small portion of the 
larger ring polymer. Within this region, the original knot 
configuration can assume a hierarchy of contracted shapes, the 
dominating one  given by just one small loop. This hierarchy 
is investigated in detail for the flat trefoil knot, and 
corroborated by Monte Carlo simulations.
\end{abstract} 
 
 
\maketitle 
 
The static and dynamic behaviour of single polymer chains, such as DNA, 
and multichain systems like gels and rubbers, is strongly influenced by 
knots and permanent entanglements \cite{deG79,DE86,Kau93,GK94}. 
Topological constraints are created with probability one during the 
polymerisation of long closed chains \cite{FWD}; more generally, knots 
and entanglements are a ubiquitous element of higher molecular 
multi-chain melts and solutions. This has profound consequences,
reaching far into biology and chemistry. For instance, knots in DNA impede 
the separation of the two strands of the double helix during transcription, 
and therefore the access to the genetic code \cite{A94}. Chemically, even 
single closed polymers may exhibit quite different properties if they 
have different topology \cite{BS99}. In the nanosciences, recent 
experimental techniques allow {\em single\/} polymer molecules
(with fixed topology) to be probed and manipulated; e.g., by single 
molecule spectroscopy \cite{MO99}, or by optomicroscopical 
imaging of small latex beads attached to a molecule \cite{micro}.
These tools provide impetus for the theoretical understanding of the 
behaviour of macromolecules under topological constraints. 
However, analytical studies, such as the statistical mechanics
of a knotted polymer, are difficult since topological constraints
require knowledge of the complete shape of the curve. Such global
constraints are hard to implement, and a complete statistical mechanical 
description of knots remains unattained \cite{KV98,VO98,Gro2000}.

\begin{widetext} 
 
\begin{figure} 
\unitlength=1cm 
\begin{picture}(6,4.4) 
\put(-7.8,-20.3){\includegraphics{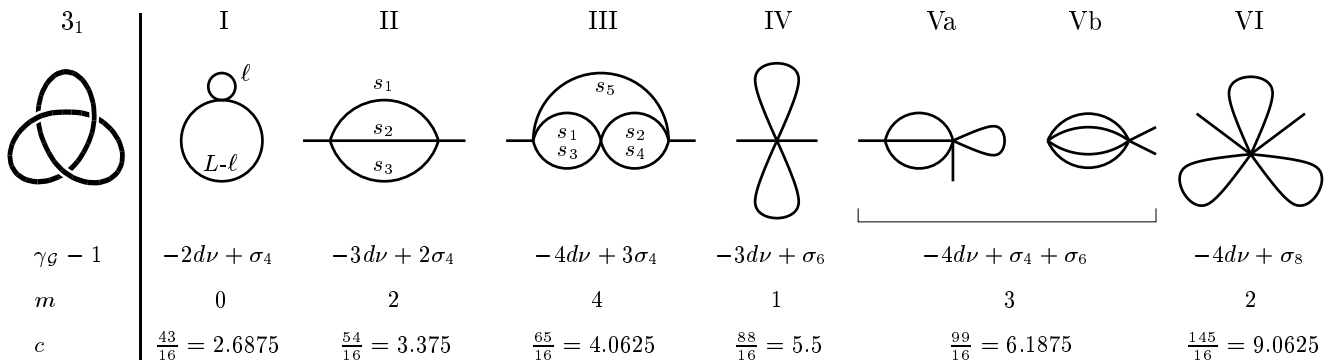}} 
\end{picture} 
\caption{{\small Standard minimal projection of the trefoil knot $3_1$,
followed by its different possible contractions,
arranged according to higher scaling orders. The uncontracted trefoil 
geometry is found at position III of the hierarchy. At I, the figure-eight 
structure is drawn. The diagrams II-VI show the multiply connected knot 
region of total length $\ell=\sum_{i=1}^{{\cal N}-1} s_i$ where the 
protruding legs indicate the outgoing large loop of length 
$s_{\cal N} = L-\ell$. Below the individual contractions, we include 
the network exponents $\gamma_{\cal G}$, the number $m$ of independent 
integrations, and the exponents $c$ defined via the PDF, 
$p(\ell) \sim \ell^{-c}$.}
\label{fig_zoo}}
\end{figure} 
 
\end{widetext}

The mathematical discipline of knot theory provides tools for the 
classification of knots. In particular, different knots can be 
distinguished by their {\em projections\/} onto a 2D plane, keeping
track of crossings according to which segment passes on top of another
\cite{Kau93,GK94}. 
By a sequence of so-called Reidemeister moves \cite{Kau93}, which leave the
topology unchanged, the number of crossings can be reduced to a minimum,
which is a simple topological invariant \cite{Kau93,GK94}. For instance,
in Fig.\,\ref{fig_zoo} we depict the minimal projection of the trefoil
knot, classified as $3_1$, with its 3 crossings.
Such quasi 2D projections, which we call {\em flat knots\/}, can be
physically realized by compressing originally 3D knots by forces 
normal to the projection plane. 
Examples include polymers adsorbed on a surface or membrane by 
electrostatic or other adhesive forces \cite{MR99}; 
or confined between parallel walls. 
In these cases the flat polymer knot can still equilibrate in
2D. Another experimental realization comes from 
Ref.\,\cite{BDVE2001}, in which macroscopic knotted chains are 
flattened by gravity onto a vibrating plane. 
The equilibrium shapes, and their scaling properties,
of such flat knots are studied in this paper. Flat knots have the
additional advantage of being easy to image by microscopy. They are
also more amenable to numeric studies than their 3D counterparts, 
and have in fact been already studied in Ref.\ \cite{GO99}.

There is growing numerical evidence that prime knots are {\em tight\/} 
in the sense that the topologically entangled region is statistically
likely to be localised on a small portion of the longer chain
\cite{GO99,KOV2000,OTJ98}. Indirect
numerical evidence of this was originally obtained by simulations 
indicating that the radius of gyration of a long polygon in 3D is
asymptotically independent of its knot type; while the presence of the knot
increases the number of configurations by a factor related
to the number of positions of the tight region around the remaining
loop \cite{OTJ98}. Simulations of 2D polygons in Ref.\,\cite{GO99} 
provide quite convincing visual evidence of localised knot regions.
In this paper, we {\em quantify the tightness of flat knots\/}, using scaling 
arguments to obtain the power law size distributions for a hierarchy
of possible equilibrium shapes. For the trefoil, Fig.\,\ref{fig_zoo}
shows this hierarchy of shapes and the corresponding exponents for
the distribution of knot size.

To demonstrate the entropic origin of tight shapes, we initially consider a
simple ring of length $L$ with one crossing. The effect of the crossing
consists in creating two loops of lengths $\ell$ and $L-\ell$, respectively,
while the orientation of the crossing is irrelevant. In this sense, the

\noindent 
crossing can be replaced with a 
vertex with four outgoing legs. We denote the resulting network by 
${\cal G}_{\text{I}}$ (see position I in Fig.\,\ref{fig_zoo}). 
In fact, we can more generally consider a sliding ring, 
or slip-link \cite{DEB}, to force two points of the chain to be close to
each other, in $d$ dimensions. Without self-avoiding 
constraints (ideal chains), the number of configurations 
$\omega_{\text{I}}(\ell,L)$ scales as \cite{deG79,rem1} 
\begin{equation} \label{oI}
\omega_{\text{I}}(\ell,L) \sim \mu^L \ell^{-d/2}(L-\ell)^{-d/2} \, , 
\end{equation} 
where, on a lattice, $\mu$ is the effective connectivity constant for 
Gaussian random walks. The average loop size is given by
$\langle\ell\rangle=\int_{a}^{L-a}d\ell\ell\omega_{\text{I}}(\ell,L)\Big/ 
\int_{a}^{L-a}d\ell\omega_{\text{I}}(\ell,L)$,
where $a$ is a short-distance cutoff set by the lattice constant.
Note that $\langle\ell\rangle=L/2$ due to symmetry.
However, the corresponding probability density function (PDF)
is strongly peaked at $\ell = 0$ and $\ell = L$, and a {\em typical\/}
shape  consists of one tight and one large loop. In $d=2$, the mean
size of the smaller loop, $\langle\ell\rangle_< \sim L/|\ln(a/L)|$,
is still rather large. It is instructive to compare to higher 
dimensions: one has {\em weak localisation}, 
$\langle\ell\rangle_< \sim a^{1/2}L^{1/2}$, in $d=3$,
and {\em strong localisation}, $\langle\ell \rangle_<\sim a$,
in $d>4$. Thus, for ideal chains, tightness of the smaller loop is more
pronounced in higher dimensions. 
 
To include self-avoiding interactions, we 
use results for general polymer networks obtained by Duplantier \cite{Dup86}, 
and in Refs. \cite{OB88,SFLD92}: In a network ${\cal G}$ consisting of 
${\cal N}$ chain segments of lengths $s_1,\ldots,s_{\cal N}$ and total
length $L=\sum_{i=1}^{\cal N}s_i$, the number of configurations 
$\omega_{\cal G}$ scales as \cite{rem2}
\begin{equation} 
\label{network} 
\omega_{\cal G}(s_1,\ldots,s_{\cal N})=\mu^{L}s_{\cal N}^{\gamma_{\cal G}-1} 
{\cal Y}_{\cal G}\left( \frac{s_1}{s_{\cal N}},\ldots,\frac{s_{{\cal N}-1}}{s_ 
{\cal N}}\right), 
\end{equation} 
where ${\cal Y}_{\cal G}$ is a scaling function, and $\mu$ is the effective 
connectivity constant for self-avoiding walks. The exponent 
$\gamma_{\cal G}$ is given by $\gamma_{\cal G}=1-d\nu{\cal L}+\sum\limits_{N
\ge 1}n_N\sigma_N$, where $\nu$ is the swelling exponent, ${\cal L}$ 
is the number
of independent loops, $n_N$ is the number of vertices with $N$ outgoing legs,
and $\sigma_N$ is an exponent associated with such a vertex.
In $d=2$, $\sigma_N = (2-N)(9N+2)/64$ \cite{Dup86}.

The network ${\cal G}_{\text{I}}$ corresponds to the parameters ${\cal N}=2$,
${\cal L}=2$, $n_4=1$, $s_1=\ell$, and $s_2=L-\ell$. By virtue of Eq.\
(\ref{network}), the number of configurations of ${\cal G}_{\text{I}}$ with
fixed $\ell$ follows the scaling form 
\begin{equation} 
\label{t_scaling} 
\omega_{\text{I}}(\ell,L)=\mu^L (L-\ell)^{\gamma_{\text{I}}-1} 
{\cal X}\left(\textstyle{\frac{\ell}{L-\ell}}\right), 
\end{equation} 
where $\gamma_{\text{I}}=1-2d\nu+\sigma_4$. In the limit $\ell\ll L$,
$\omega_{\text{I}}(\ell,L)$ should reduce to the number 
$\omega_{\text{crw}}(L) \sim \mu^L L^{-d \nu}$
of {\em closed random walks\/} of length $L$ which start 
and end at a given point in space \cite{rem1,KMP2000}.
This implies ${\cal X}(x)\sim x^{\gamma_{\text{I}}-1+d\nu}$
as $x \to 0$, such that 
\begin{equation} 
\label{sl_limit} 
\omega_{\text{I}}(\ell,L)\sim\mu^L (L-\ell)^{-d\nu}\ell^{-c} \, , 
\quad \ell\ll L \, , 
\end{equation} 
where $c=-(\gamma_{\text{I}}-1+d\nu)=d\nu-\sigma_4$. Using $\sigma_4=-19/16$
and $\nu = 3/4$ in $d=2$, we find $c=43/16=2.6875$. In $d = 3$,
$\sigma_4\approx -0.48$ and $\nu\approx 0.588$, so that $c\approx 2.24$
\cite{SFLD92,KMP2000}. 
In both cases the result $c>2$ implies that the loop of length 
$\ell$ is strongly localised in the sense defined above. 
This justifies the a priori assumption $\ell\ll L$, 
and makes the analysis self-consistent. 
Note that for self-avoiding chains, in $d=2$ the 
localisation is even {\em stronger\/} than in $d=3$, in contrast 
to the corresponding trend for ideal chains. 

\begin{figure}
\unitlength=1cm
\begin{picture}(8,6)
\put(-2.2,-18.2){\includegraphics{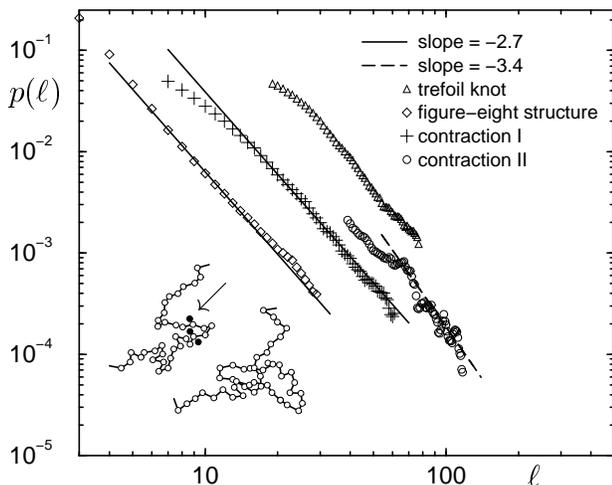}}
\end{picture}
\caption{{\small
Power law tails in PDFs for the size $\ell$ of tight segments: 
As defined in the figure, we show results for the smaller loop in a
figure-eight structure, the overall size of the trefoil knot,
as well as the two leading contractions of the latter.
The insets show typical configurations of the small loop for a 
figure-eight (the arrow points to the slip-link
consisting of 3 tethered beads), and the trefoil.
\label{figsl1}}}
\end{figure}

We performed Monte Carlo (MC) simulations (see below for details) of a
chain in  $d=2$ confined to the ${\cal G}_I$ (figure-eight) structure 
by a slip-link. As shown in Fig.\ \ref{figsl1}, the size distribution
for the small loop can be fitted to a power law with exponent $c=2.7
\pm 0.1$ \cite{REMM}, in good agreement to the above prediction.
  
For the configuration ${\cal G}_I$, the probability for the size of 
each loop is
peaked at $\ell\to 0$ and $\ell\to L$. For more complicated projections, the
joint probability to find the individual segments with given lengths $s_i$
is expected to peak at the edges of the higher-dimensional configuration
hyperspace. Some analysis is necessary to find the optimal shapes;
as presented here for the simplest non-trivial knot, the (flat) trefoil 
knot $3_1$ (see Fig.\,\ref{fig_zoo}).
Each of the three crossings is replaced with a vertex with four 
outgoing legs, and the resulting network is assumed to separate into 
a large loop and a multiply connected region which includes the vertices. 
Let $\ell=\sum_{i=1}^5 s_i$ be the total 
length of all segments contained in the
multiply connected knot region (see Fig.\,\ref{fig_zoo}, position III). 
Accordingly, the length of the large loop is $L-\ell$. In the limit 
$\ell\ll L$, the number of configurations of the network 
${\cal G}_{\text{III}}$ can be derived in a similar way as above,  
yielding 
\begin{equation} 
\label{ts_limit} 
\omega_{\text{III}}'\sim\mu^L (L-\ell)^{-d\nu}\ell^{\gamma_{\text{III}}-1+d\nu} 
{\cal W}\left(\frac{s_1}{\ell},\frac{s_2}{\ell},\frac{s_3}{\ell}, 
\frac{s_4}{\ell}\right), 
\end{equation} 
where $\gamma_{\text{III}}=1-4d\nu+3\sigma_4$ and ${\cal W}$ is a 
scaling function. The prime on $\omega_{\text{III}}$ indicates that each of 
the segment lengths $s_i$ is fixed. In order to obtain the number of 
configurations $\omega_{\text{III}}(\ell,L)$ for the case we are interested 
in, where only the total length $\ell$ is fixed, we should integrate $\omega_ 
{\text{III}}'$ over all distributions of lengths $s_i$
under the constraint $\sum_{i=1}^{5} s_i = \ell$. This leads to the result 
\begin{equation} 
\label{t_limit} 
\omega_{\text{III}}(\ell,L)\sim\mu^L (L-\ell)^{-d \nu}\ell^{-c},
\end{equation} 
with $c=-(\gamma_{\text{III}}-1+d\nu)-m$, where $m = 4$ corresponds to the 
number of independent integrations over $s_i$. Thus, $c=3d\nu-3\sigma_4-4= 
\frac{65}{16}$ (see Fig.\,\ref{fig_zoo}, position III). 

However, some care is necessary in performing these integrations, 
since the scaling function ${\cal W}$ in Eq.\,(\ref{ts_limit}) may 
exhibit non-integrable singularities if one or more of its arguments 
tend to $0$ or $1$. The geometries corresponding to these 
limits (edges of the configuration hyperspace)
represent {\em contractions\/} of the original 
trefoil network ${\cal G}_{\text{III}}$ in the sense that the length of 
one or more of the segments $s_i$ is of the order of the 
short-distance cutoff $a$. If such a short segment connects 
different vertices, they cannot be resolved on larger length 
scales, but melt into a single, new vertex, in the 
context of our scaling analysis \cite{ZJ89}. Thus, each 
contraction corresponds to a different network ${\cal G}$,
which may contain a vertex with six or even eight outgoing legs. 
For each of these networks, one can calculate the corresponding 
exponent $c$ in a similar way as above, and 
using the Euler relations
$2 {\cal N} = \sum_{N \ge 1} N n_N$ and
${\cal L} = \sum_{N \ge 1} \frac{1}{2} (N-2) n_N + 1$,
we obtain
\begin{equation}
\label{euler}
c_{\cal G}=2+\textstyle{\sum_{N\ge 4}n_N\left[\frac{N}{2}(d\nu-1)+(|\sigma_N|-
d\nu)\right]}.
\end{equation}

Our scaling analysis relies on an expansion in 
$a/\ell\ll 1$, and the values of $c$ determine a sequence of 
contractions according to higher orders in $a/\ell$: The {\em smallest\/} 
value of $c$ corresponds to the most likely contraction, while the  
others represent corrections to this leading scaling behaviour, and
are thus less and less probable (see Fig.\ \ref{fig_zoo}). 
To lowest order, the trefoil behaves like a large ring polymer at whose fringe
the point-like knot region is located. At the next level of resolution,
it appears contracted to the figure-eight shape 
(see Fig.\,\ref{fig_zoo}, position I).
If more accurate data are available, the higher order
shapes II to VII may be found with decreasing 
probability. Interestingly, the original uncontracted trefoil 
configuration ranks third in the hierarchy of shapes. 
Note that the contractions shown in Fig.\ \ref{fig_zoo} may
occur in different topological variants. For instance,
the smaller loop in contraction I could be inside the larger loop. However,
this does not make a difference in terms of the scaling analysis.

To check these predictions, we performed MC simulations 
of a flat trefoil knot using a standard bead-and-tether chain model
in 3D. The polymer was flattened by a ``gravitational''
field $V=-k_BTh/h^*$ perpendicular to a hard wall, where $h$ is the 
height and $h^*$ was set to $0.3$ times the bead diameter. 
With 512 beads, around $10^{11}$ MC steps were performed to generate
equilibrium states. Configurations corresponding to contraction I
are then selected by requiring that besides a large loop, they contain
only one segment larger than a preset cutoff length (taken to be 5 monomers),
and similarly for contraction II.
The size distributions for such contractions, as well as for all
possible knot shapes are shown in Fig.\ \ref{figsl1}. 
The tails of the distributions are indeed consistent with the predicted
power laws, although the data (especially for contraction II) is too
noisy for a definitive statement.

These scaling results pertain also to other prime knots.
In particular, the dominating contribution for {\em any} prime knot 
corresponds to the figure-eight contraction ${\cal G}_{\text{I}}$. 
This can be shown by noting that Eq.\ (\ref{euler}) predicts a
larger

\begin{widetext}

\begin{figure} 
\unitlength=1cm 
\begin{picture}(20,3.8) 
\put(-2,-22.3){\includegraphics{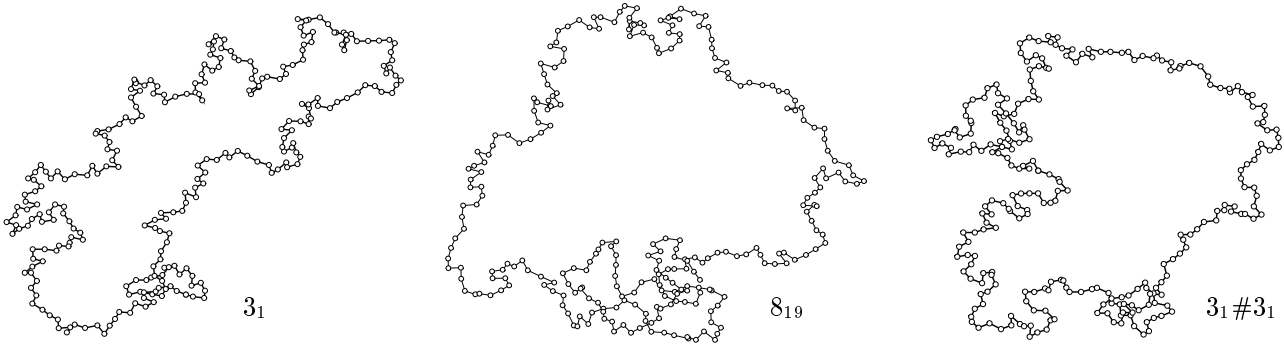}} 
\end{picture} 
\caption{{\small Typical equilibrium configurations for
the trefoil $3_1$,
the prime knot $8_{19}$, and the composite knot $3_1$\#$3_1$
consisting of two trefoils, in $d=2$. The initial conditions were symmetric.}
\label{fig_mcknots}} 
\end{figure} 
 
\end{widetext}

\noindent
value of the scaling exponent $c$ for any network ${\cal G}$ other
than ${\cal G}_{\text{I}}$.
Figure \ref{fig_mcknots} demonstrates the tightness of the prime
knot $8_{19}$. Composite knots, however, can maximise the number 
of configurations by splitting into their prime factors as indicated
in Fig.\,\ref{fig_mcknots} for $3_1$\#$3_1$. Each prime factor is 
tight and located at the fringe of one large loop, and accounts for 
an additional factor of $L$ for the number of configurations, as 
compared to a ring of length $L$ without a knot. 
Indeed, this gain in entropy leads to the tightness of knots.

In conclusion, we find that the trefoil knot, as well as higher order 
prime and composite knots, are sharply localized when forced to lie flat.
In the most likely shapes, each prime factor is tightened into a
loop (a figure-eight contraction). 
It is natural to speculate that entropic factors also confine knots in 
$d=3$ by power law distributions in size.
Direct checks of such behaviour are hampered by the difficulty of
identifying the knotted region of a curve \cite{KOV2000} in $d=3$.
One may instead search for indirect signatures of localized knots
in detailed dependencies of gyration radius and other polymeric
quantities on length.

RM and AH acknowledge financial support from the DFG. PGD acknowledges
financial support from the Research Council of Norway. Support from
the NSF (DMR-01-18213 and PHY99-07949) and the US-Israel BSF (1999-007)
is also acknowledged.


\vspace{-2mm}

{\small 
 
\begin{thebibliography}{99} 
 
\bibitem{deG79}P.-G. de Gennes, {\em Scaling concepts in polymer 
physics} (Cornell University Press, Ithaca, New York, 1979). 
 
\bibitem{DE86}M. Doi and S.F. Edwards {\em The Theory of 
Polymer Dynamics} (Clarendon Press, Oxford, 1986). 
 
\bibitem{Kau93}L.H. Kauffman, {\em Knots and physics} 
(World Scientific, Singapore, 1993). 
 
\bibitem{GK94}A.Yu. Grosberg and A.R. Khokhlov, {\em Statistical 
Mechanics of Macromolecules} (AIP Press, New York, 1994). 
 
\bibitem{FWD} 
H.L. Frisch and E. Wasserman,  
J. Am. Chem. Soc. {\bf 83}, 3789 (1961); M. Delbr\"uck,
Proc. Symp. Appl. Math. {\bf 14}, 55 (1962). 
 
\bibitem{A94}B. Alberts et al., {\em The molecular biology of the 
cell} (Garland, New York, 1994). 
 
\bibitem{BS99} J.-P. Sauvage and C. Dietrich-Buchecker, 
{\em Molecular catenanes, rotaxanes, and knots: a journey 
through the world of molecular topology\/} 
(Wiley-VCH, Weinheim, 1999). 
 
\bibitem{MO99}W.E. Moerner and M. Orrit, Science {\bf 283}, 1670 (1999). 
W.E. Moerner and L. Kador, Phys. Rev. Lett. {\bf 62}, 2535 (1989). 
  
\bibitem{micro}A. van Oudenaarden and J.A. Theriot, Nat. Cell. Biol. {\bf 1}, 
493 (1999); J.A. Theriot et al., Nature {\bf 37}, 257 (1992). 
 
\bibitem{VO98}T.A. Vilgis and M. Ott, Phys. Rev. Lett. {\bf 80},  
881 (1998). 
 
\bibitem{Gro2000}A.Yu. Grosberg, Phys. Rev. Lett. {\bf 85},  
3858 (2000). 
 
\bibitem{KV98} A.L. Kholodenko and T.A. Vilgis, Phys. Rep. {\bf 298}, 
251 (1998). 
 
\bibitem{MR99}B. Maier and J.O. R\"adler, Phys. Rev. Lett. {\bf 82},  
1911 (1999). 

\bibitem{BDVE2001}E. Ben-Naim et al., Phys. Rev. Lett. {\bf 86}, 2001. 
(Whether such shaking results in equilibrium states is
not clear.)
 
\bibitem{GO99}E. Guitter and E. Orlandini, J. Phys. A {\bf 32}, 
1359 (1999). 
 
\bibitem{KOV2000}V. Katritch et al., Phys. Rev. E {\bf 61}, 5545 (2000). 
 
\bibitem{OTJ98}E. Orlandini, M.C. Tesi, E.J. Janse van Rensburg,
and S.G. Whittington, J. Phys. A {\bf 31}, 5953 (1998). 
 
\bibitem{DEB}M. Doi and S.F. Edwards, J. Chem. Soc. Faraday Trans. 
II {\bf 74}, 1802 (1978); R.C. Ball et al., Polymer {\bf 22}, 1010 (1981). 
 
\bibitem{rem1}Following Refs.\,\cite{OTJ98,GO99}, we consider two 
configurations as distinct if they cannot be superimposed by translation.
This  eliminates the degrees of freedom due to translational 
invariance of the whole structure.

\bibitem{Dup86}B. Duplantier, Phys. Rev. Lett. {\bf 57}, 941 (1986); 
J. Stat. Phys. {\bf 54}, 581 (1989). 
 
\bibitem{SFLD92}L. Sch\"afer et al., Nucl. Phys. B {\bf 374}, 473 (1992). 
 
\bibitem{OB88}K. Ohno and K. Binder, J. Phys. (Paris) {\bf 49}, 1329 (1988). 
 
\bibitem{rem2}This result is valid if the network has at least one
vertex with $N \neq 2$ outgoing legs. In contrast, for a simple ring 
polymer of length $L$, one has 
$\omega(L) = \mu^{L} L^{- d \nu - 1}$ \cite{rem1}.

\bibitem{KMP2000}Y. Kafri et al., Phys. Rev. Lett. {\bf 85}, 4988 (2000); 
preprint cond-mat/0108323 (2001). 

\bibitem{REMM} The quoted errors reflect our subjective estimate of possible
systematic errors.
 
\bibitem{ZJ89}In field theory, this is an example of an 
operator product expansion; 
see J. Zinn-Justin, {\em Quantum Field Theory and 
Critical Phenomena\/} (Clarendon Press, Oxford, 1989). 
 
\end{thebibliography} 
}
\end{document}